\documentclass[fleqn,10pt]{wlscirep}
\usepackage[utf8]{inputenc}
\usepackage[T1]{fontenc}
\title{Role of Motility and Nutrient Availability in Drying Patterns of Algal Droplets}

\author[1,2 *]{Anusuya Pal}
\author[3]{Anupam Sengupta}
\author[4, 5, 6]{Miho Yanagisawa}
\affil[1]{Department of Physics, University of Warwick, Coventry, CV47AL, UK}
\affil[2]{Graduate School of Arts and Sciences, The University of Tokyo, Komaba 4-6-1, Meguro, Tokyo, 153-8505, Japan}
\affil[3]{Physics of Living Matter Group, Department of Physics and Materials Science, University of Luxembourg, L-1511, Luxembourg}
\affil[4]{Komaba Institute for Science, Graduate School of Arts and Sciences, The University of Tokyo, Komaba 3-8-1, Meguro, Tokyo 153-8902, Japan}
\affil[5]{Graduate School of Science, The University of Tokyo, Hongo 7-3-1, Bunkyo, Tokyo 113-0033, Japan}
\affil[6]{Center for Complex Systems Biology, Universal Biology Institute, The University of Tokyo, Komaba 3-8-1, Meguro, Tokyo 153-8902, Japan}

\affil[*]{apal@g.ecc.u-tokyo.ac.jp}


\begin{abstract}
Sessile drying droplets in various bio-relevant systems, encompassing passive bio-colloids like DNA, proteins, and blood to active microbes, gain considerable attention due to intricate interplay among different convective flows, droplet pinning, mechanical stress, wettability, and the emergence of distinctive patterns. \textit{Chlamydomonas reinhardtii}, or chlamys, is a versatile algal model employed in molecular biology research and spanning diverse biotechnological realms. While chlamys are harnessed at single-cell and population levels, their exploration in the context of drying sessile droplets remains limited. This paper illuminates the multifaceted potential of chlamys, delving into motility-nutrient interactions and their role in emergent morphological patterns. The interplay of two competing stressors--localized nutrient scarcity and mechanical stress during drying--is investigated. Irrespective of these stressors, the global mechanical stress fails to induce any cracks during the drying process. Interestingly, the reverse ``coffee-ring effect" is predominantly observed in the non-motile chlamys in the presence of local nutrients whereas the nutrient depletion prompts local stress in motile chlamys, culminating in cooperative aggregation and cluster formation. Furthermore, the quantitative image processing technique leverages textural statistics to classify the patterns into four classes, motile+with nutrients, motile+without nutrients, non-motile+with nutrients, and non-motile+without nutrients, with five distinct drying stages-- Droplet Deposition, Capillary Flow, Dynamic Droplet Phase, Aggregation Phase, and Dried Morphology. 
\end{abstract}
\begin{document}

\flushbottom
\maketitle
%
%
\thispagestyle{empty}

\section{Introduction}
\label{sec:intro}

Microbial active matter \cite{sengupta2020front} encompasses a diverse microbial species, including bacteria, algae, fungi, bacteriophages, and viruses, which are ubiquitously present in nature. These systems have been extensively studied at single-cell level \cite{drescher2010direct}, and population level \cite{Bittermann2021light}, serving as subjects of fundamental research and various biotechnological applications. Their unique intrinsic traits, such as size, shape, structure, growth characteristics, and responses to external stimuli like temperature, humidity, stress, pH, and nutrient availability \cite{PAL2023102870}, have contributed to their scientific significance.

The burgeoning field of single-cell biology (mostly in the nano to microscales) has been propelled by innovative tools like micro-arrays, facilitating investigations into cellular behavior at solid-liquid interfaces. Techniques like microfluidics \cite{juang2016applications} and evaporative drying of droplets \cite{Thokchom2014, Sempels2013, Bittermann2021light} have been harnessed to explore microorganism dynamics on a population level. The process of evaporative drying of biologically relevant systems induces convective flow, resulting in collective behavior, self-assembly, mechanical instabilities, and the emergence of macroscopic patterns \cite{PAL2023102870}.

Drying of sessile \textit{active droplets}-- droplet confining active inclusions like bacteria \cite{Thokchom2014, kang2020simple, Majee2021}, bacteriophages \cite{Huang2021}, spermatozoa \cite{rios2018pattern}, nematodes \cite{peshkov2022synchronized}, and so on--have received considerable attention in recent years \cite{araujo2023sm}. The drying droplets containing active biological particles are notably more intricate than that passive particles due to the added complexities of motility and micro-environmental influences on fluid flows, in addition to the self-aggregation common in bio-colloidal droplets. Intriguingly, the ``coffee-ring" effect \cite{deegan1997capillary} can be altered by bio-surfactants produced by bacteria \cite{Sempels2013}, and various factors, including substrate properties, bacterial strains \cite{susarrey2016pattern, susarrey2018bacterial}, particle characteristics (size, shape), and drying rates \cite{Richard2020} impact the resulting deposited patterns. However, it is not limited to only active systems but also includes passive systems such as DNA and nucleic acids \cite{ye2022evaporative}, polymers \cite{fujisawa2023molecular, liu2023three, bhardwaj2010self, wang2022wetting}, liquid crystal \cite{pal2019comparative, pal2022hierarchical}, blood \cite{pal2020concentration}, and so on. Within the domain of sessile drying droplet studies, an exciting avenue has emerged in the realm of disease diagnosis. The distinctive patterns that emerge during the drying process not only have the potential to differentiate between various diseases but also offer insights into the disease's developmental stage \cite{PAL2023102870}.

Recent advancements in biomaterials and tissue engineering have underscored the remarkable potential of \textit{Chlamydomonas reinhardtii}, (henceforth chlamys), as a versatile candidate for various applications. It spans from growth factor delivery in wound healing scaffolds to serving as an oxygen source in scenarios of respiratory failure or organ transplantation \cite{schenck2015photosynthetic}. Demonstrating remarkable adaptability to even hostile micro-environments, chlamys exhibits tailored responses contingent upon stressors such as oxygen levels, light intensity, nutrient availability, and toxins \cite{de2019unity}. What further enhances its appeal is its optimal size, mechanosensitivity, and capacity for biocompatible cell wall modification without releasing toxins upon decomposition, rendering it a superior choice for drug delivery compared to other flagellated microbes \cite{doi:10.1021/acsbiomaterials.2c01389}. Notably, these chlamys elucidate the molecular underpinnings of specific human ciliary diseases, including ciliary dyskinesia and ailments linked to cystic kidneys \cite{wijffels2013potential}. This newfound focus on chlamys has unveiled insights into the intricate mechanisms underlying these diseases, shedding light on potential therapeutic avenues.

Despite being extensively studied at the population level (mostly in the micro to macroscales), the investigation of chlamys in sessile drying droplet settings remains limited. Only one study has explored chlamys drying droplets, analyzing the impact of light stimuli on its responses and resultant patterns-- linked to motility \cite{Bittermann2021light}. However, a crucial determinant, nutrient conditions (beyond motility), remains unexplored. Across diverse species, nutrient availability is known to impact motility and vice-versa \cite{sengupta2022scienceadv}, yet the coupling between motility, physiology, and pattern formation in drying algal droplets remains uncharted territory.

\begin{figure*}[h]
\centering
  \includegraphics[height=12cm]{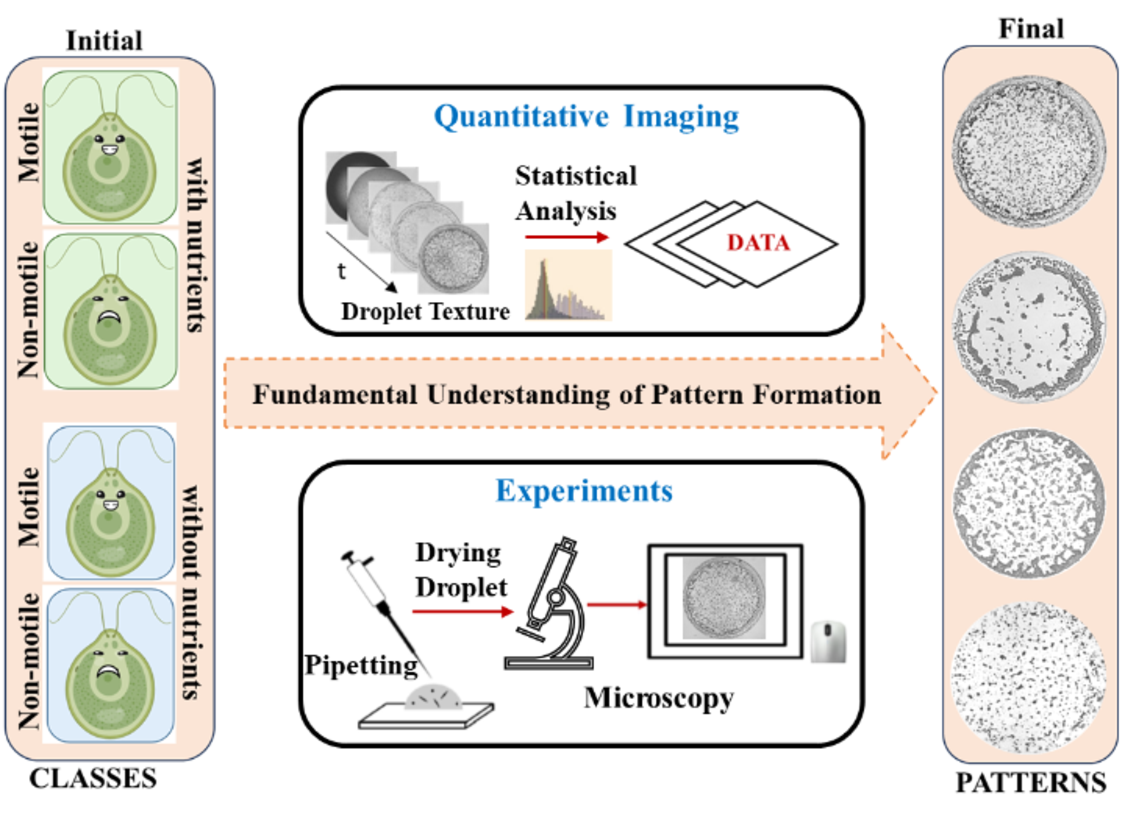}
  \caption{It explores the drying behavior of four distinct \textit{Chlamydomonas reinhardtii} algal droplets. It unfolds the motility-nutrient effects on the morphological patterns by using simple drying droplet experiment and the quantitative image textural statistics.}
  \label{fig1flow}
\end{figure*}

To address this research gap, this paper presents an experimental study with a quantitative image analysis of the drying process and pattern formation within chlamys droplets. In this context, the drying process functions as an engine, propelling the droplet (system) from one equilibrium state to another. It seeks to answer two fundamental questions: (i) what is the role of motility in pattern dynamics during the drying process? and (ii) Is there any correlation between the motility and the local nutrient availability deciding the drying patterns? If so, how?   

Figure~\ref{fig1flow} outlines a schematic flowchart depicting the approach of this study. It addresses the fundamental understanding of the drying behavior of chlamys droplets by exploiting simple drying process and studying the pattern formation. The two aspects are examined, one is the effects of the motility and the local nutrient availability, and second is the two competing stress factors: one stemming from inadequate local nutrient availability and the other originating from substantial water loss during the drying process. This study also identifies distinct drying stages using the quantitative image textual statistical features (mean, standard deviation, kurtosis, and skewness), and assess the ``coffee-ring" effect across the four specified classes (motile+with nutrients, motile+without nutrients, non-motile+with nutrients, and non-motile+without nutrients). As these systems involve multiple factors across varying scales, the drying of chlamys droplets serves as an initial avenue to steer subsequent in-depth inquiries.

\section{Results and Discussions}
\label{sec:res}

\subsection{Visual inspection on drying evolution of chlamys droplets }
\label{subsec:time}

\begin{figure*}[h]
\centering
  \includegraphics[height=14cm]{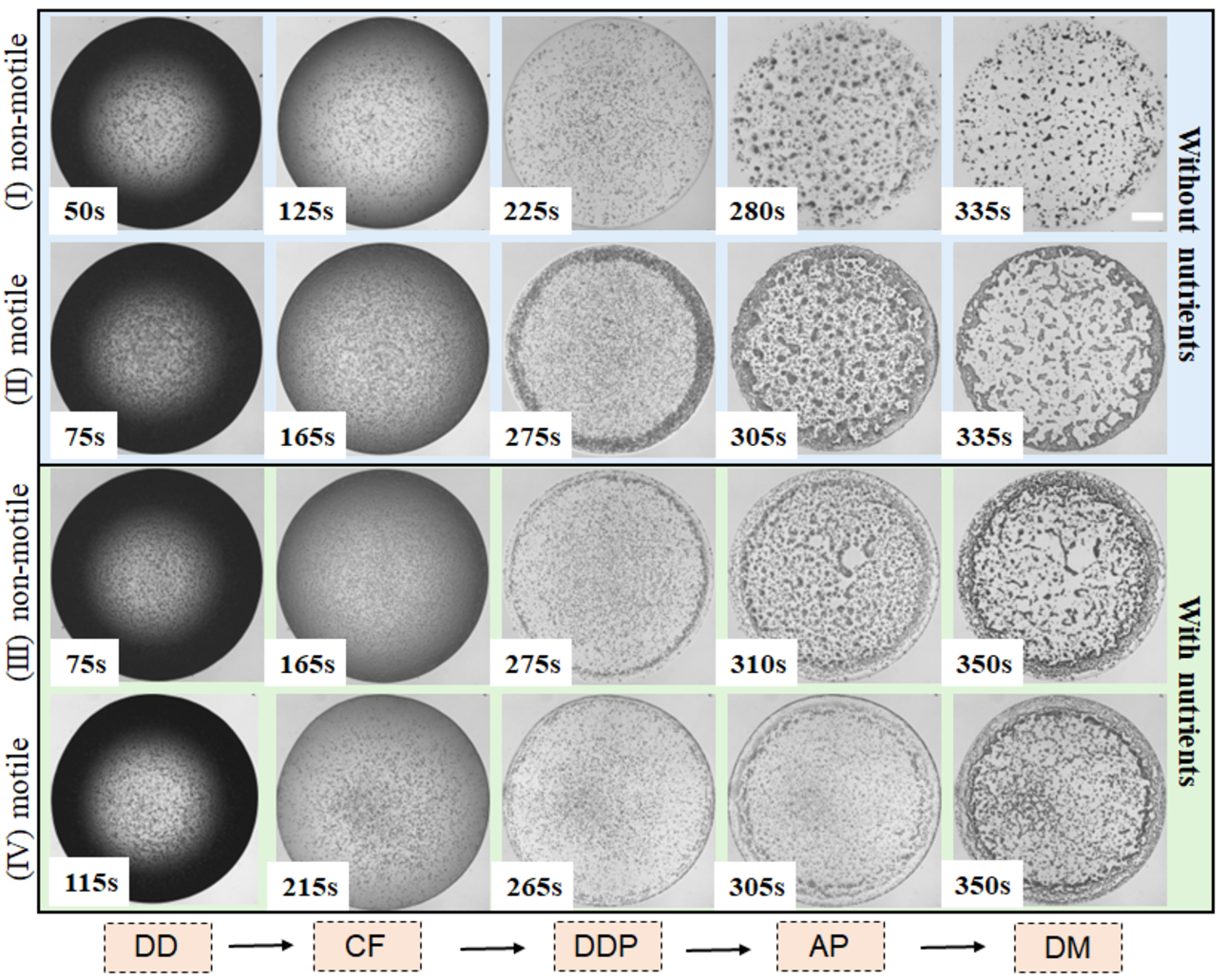}
  \caption{Drying evolution of chlamys droplets for various classes: (I) without nutrients+non-motile, (II) without nutrients+motile, (III) with nutrients+non-motile, and (IV) with nutrients+motile. The arrows show the time (t) evolution of the dynamic patterns during drying. Different stages include Droplet Deposition (DD), Capillary Flow (CF), Dynamic Droplet Phase (DDP), Aggregation Phase (AP), and Dried Morphology (DM). The scale bar of length 0.3 mm is shown with the white rectangle in the top-right panel.}
  \label{fig1a}
\end{figure*}

Figure~\ref{fig1a}(I-IV) vividly presents the evolving drying progression of both motile and non-motile chlamys droplets, with and without localized nutrient availability. The contact angle ad the height of these droplets exhibit a consistent monotonic decrease [see Figure S1 of the supplementary section], aligning with behavior observed in other bio-colloids \cite{pal2020comparative, pal2019phase}. Over time, the texture of the images transitions from dark to light, as showcased in Figure~\ref{fig1a}(I-IV). Notably, all chlamys droplets undergo four distinct stages, completing the drying process within 300-350 seconds. In Stage 1, referred to as the droplet deposition (DD) stage, the droplet adopts a spherical-cap shape upon being placed onto the coverslip (substrate). The evaporative flux is highest in the vicinity of the three-phase contact line. Importantly, the droplet size in this study (with a diameter of ${\sim}2$~mm) remains below the capillary length threshold. Consequently, surface tension prevails over gravity, preventing the droplet from flattening onto the substrate. As a result, the droplet remains pinned throughout the entirety of the drying process.

The contrast of dark black near the periphery and gray within the central region arises due to the droplet's curvature, as depicted in Figure~\ref{fig1a}(I-IV) DD stage. While internal convective flow ensues during this period, the subsequent stage (Stage 2) is marked by dominant outward capillary flow (CF), promoting the movement of chlamys toward the droplet's edge. Notably, most chlamys congregates near the droplet's periphery, generating the well-known ``coffee-ring" effect \cite{deegan1997capillary}. However, this phenomenon is less pronounced in non-motile chlamys droplets lacking local nutrients. This phase, referred to as the Dynamic Droplet Phase (DDP), is particularly distinctive for all droplet types—motile and non-motile, with and without local nutrient availability. As the drying process advances, water evaporation leads to chlamys aggregation (Figure~\ref{fig1a}(I-IV) AP Stage), a process characteristically short, requiring 10-25 seconds. In the AP stage, however, we see that the fluid-front proceeds from the periphery to the central region, promoting the self-assembly, and aggregation of chlamys in a chaotic way. The DDP and AP stages, being highly dynamic and unique, ultimately dictate the final patterns within the Dried Morphology (DM) stage (see Figure~\ref{fig1a}(I-IV) Stage DM).

\subsection{Physical mechanism of chlamys droplets}
\label{subsec:phyb}

\subsubsection{Chlamys droplets in the presence of local nutrients}
\begin{figure*}[h]
\centering
  \includegraphics[height=10cm]{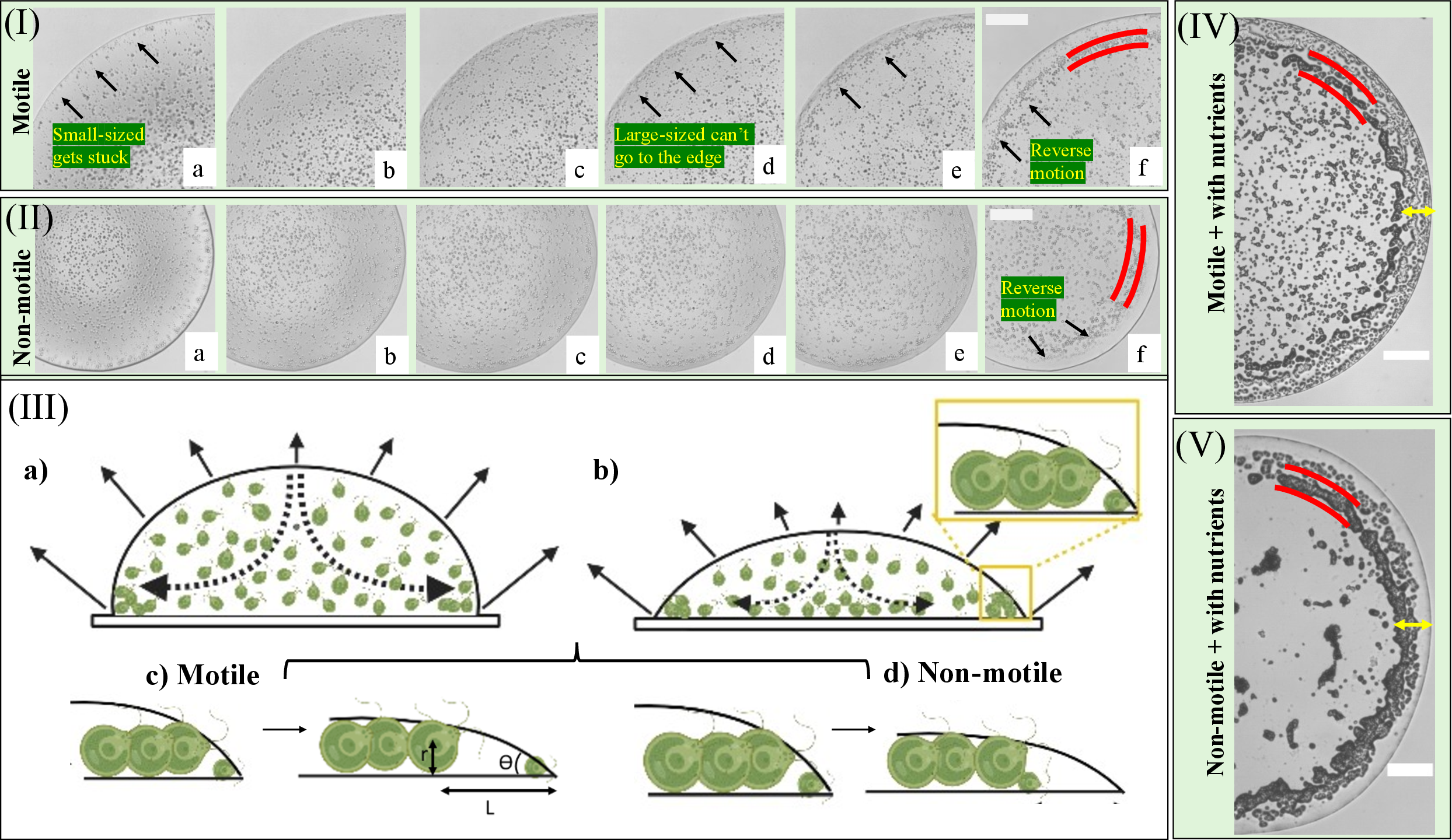}
  \caption{a-f) Time evolution of the (I) motile and (II) non-motile chlamys in the nutrient-rich environment. The images are captured under $10\times$ magnification. (III) A physical mechanism in which (a-b) Stages DD and CF in the droplet are displayed. Stages DDP is shown in c and d, with motile and non-motile chlamys, respectively. The $\theta$ represents the contact angle of the droplet, $r$ denotes the radius of the chlamys, and $L$ signifies the distance the chlamys move inward. The unique patterns in DM Stage for the (IV) motile and (V) non-motile chlamys are exhibited, where the red lines depict the deposited chlamys near the periphery, and the yellow arrows dictate the ``coffee-ring" width. A scale bar has a length of 0.2 mm. }
  \label{fig6}
\end{figure*}

Figure~\ref{fig6}(I-II)a-f provides a step-by-step illustration of the evolving drying dynamics for both motile and non-motile chlamys in the presence of localized nutrients. A visual representation of the underlying physical mechanism driving this pattern formation is depicted in Figure~\ref{fig6}(III).

For the case of motile chlamys [Figure~\ref{fig6}(I)a-f], smaller-sized chlamys (with sizes of $5-7~\mu$m) tend to accumulate near the droplet's periphery. Interestingly, larger-sized chlamys do not gather at the contact line. This phenomenon could be attributed to the significant reduction in the contact angle between the 260-300 second time interval, leading to contact angles of $12-8^{\circ}$. The relatively larger size of these chlamys prevents them from reaching the extreme edge of the droplet. As time progresses, chlamys start to aggregate near the droplet's periphery, forming a distinct ``coffee-ring" pattern \cite{deegan1997capillary}. This size-dependent phenomenon could potentially lead to segregation, as observed in previous studies \cite{liu2019segregation}.

Intriguingly, even though chlamys transport towards the periphery continues [Figure~\ref{fig6}(I-II)c-d and yellow arrows in Figure~\ref{fig6}(IV-V)], a reversal becomes evident. As water continues to evaporate, these chlamys tend to move inward (i.e., towards the droplet's central region). Remarkably, this behavior resembles the ``reverse coffee-ring" effect; however, this effect is not limited to this specific active system, as similar patterns have been observed in other drying systems \cite{weon2010capillary}. The unique aspect here is that the smaller-sized transported chlamys become entrapped while only the larger-sized counterparts move with the fluid. The histogram of different-sized chlamys is shown in Figure S2 of the supplementary section. This is in contrast to the non-motile case, where all chlamys move with the flow towards the central region [Figure~\ref{fig6}(I-II)e-f and Figure~\ref{fig6}(III)c-d]. A geometric model that supports this reverse motion can be formulated using the relationship tan~$\theta = r/L$, where $\theta$ represents the contact angle of the droplet, $r$ denotes the radius of the chlamys, and $L$ signifies the distance the chlamys move inward. With chlamys radius $r \sim 4.5~\mu m$ and inward movement distance $L \sim 25~\mu m$ (measured using ImageJ \cite{abramoff2004image}), the calculated $\theta$ is $10.2^{\circ}$. This aligns closely with the experimentally observed $\theta = 12-10^{\circ}$ obtained from contact angle measurements (as shown in Figure S1 of the supplementary section).

Two intriguing questions arise from these observations for both non-motile and motile chlamys: (a) the cause behind the pinning of these droplets and (b) the mechanism underlying the ``reverse coffee-ring" effect. Addressing the first question, it is inferred that droplet pinning is primarily driven by the presence of ions, proteins, and salts in the TAP medium. While the accumulation of small-sized chlamys might enhance pinning at the three-phase contact line, the secretion of polymer-like substances \cite{Sempels2013} from these chlamys, especially under stress \cite{devadasu2021enhanced}, is also possible. However, it is important to note that pinning is observed even in the non-motile case, where there is no possibility of secretion, as these chlamys are non-viable before the drying process begins.

Regarding the second question about the ``reverse coffee-ring" effect, one needs to understand the differences between the non-motile and motile cases. The disparity could be attributed to the fact that small chlamys experience mechanical stress as they attempt to cope with the drying process. This leads to their eventual entrapment and death, as no more fluid is available near the periphery. This phenomenon does not apply to non-motile chlamys, which behave as passive bio-colloidal particles. Consequently, when the fluid front recedes during the AP stage, it induces non-motile chlamys to migrate inward, much like other bio-colloidal droplets \cite{pal2020comparative}. This leads to a uniform texture forming in the outermost region of the droplet.

\begin{figure*}[h]
\centering
  \includegraphics[height=13.5cm]{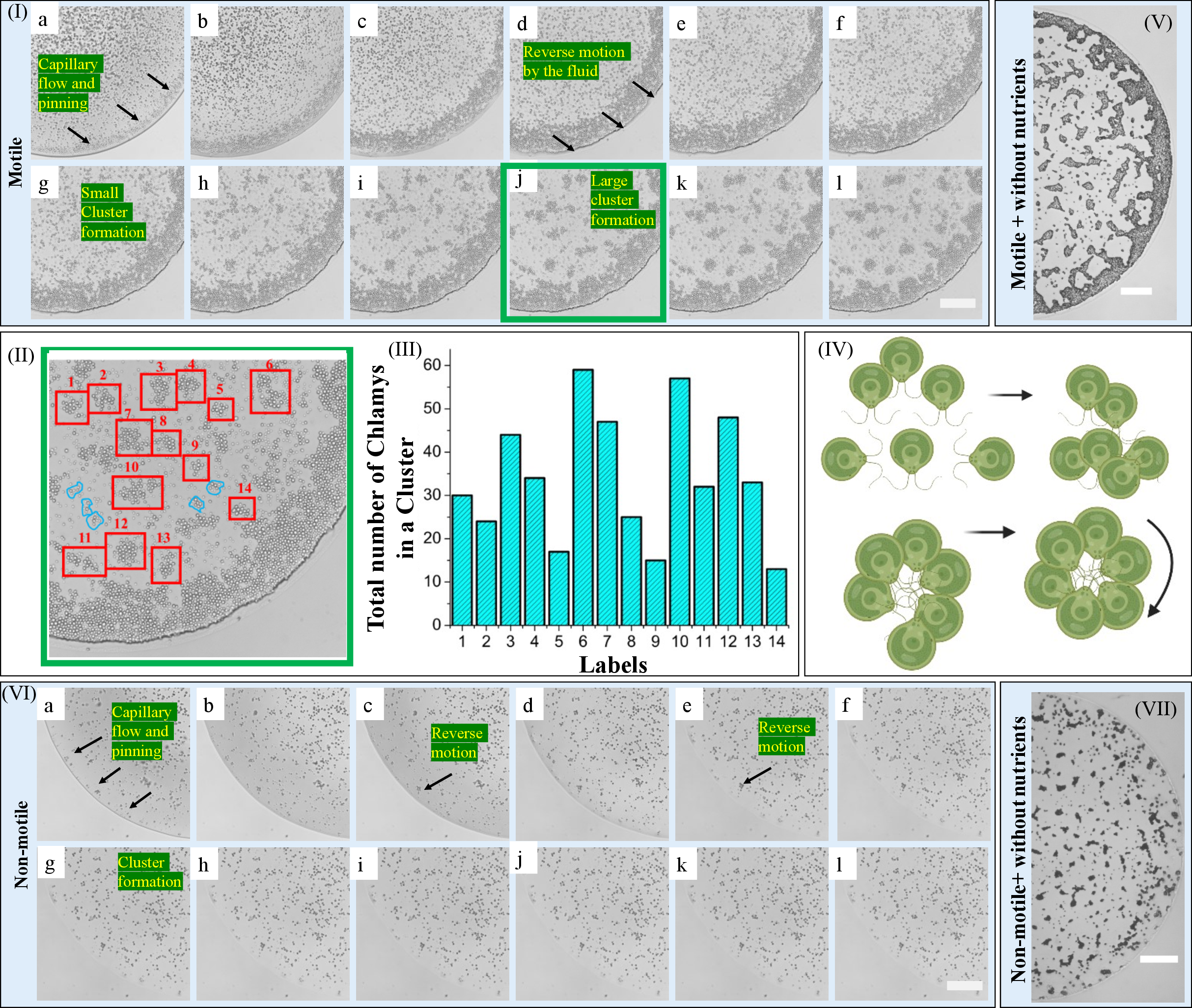}
  \caption{(I)a-l: Time evolution of the motile chlamys in the nutrient-deficient environment. The images are captured under $10\times$ magnification. (II) Zoomed view (green) of (I)j displaying clusters outlined in red with labels and small clusters in blue. (III) A histogram showing the total number of chlamys in each red-labeled cluster. (IV) A physical mechanism of cluster formation in the DDP stage is discussed. (V) Dried morphological patterns of the motile chlamys. (VI)a-l Time evolution and (VII) Dried morphological patterns of the non-motile chlamys in the nutrient-deficient environment. A scale bar has a length of 0.2 mm.}
  \label{fig7}
\end{figure*}

Importantly, the droplets under consideration do not display any signs of buckling or cracking as the drying process progresses. As the drying process nears completion, chlamys start to aggregate due to the decreasing fluid content, resulting in the formation of big aggregates or cluster-like structures. The rapid evaporation of trapped water between neighboring aggregates leads to their displacement to some extent [red lines in Figure~\ref{fig6}(I-II)f and (IV-V)]. This process culminates in Stage DM, which captures the final morphological patterns as the drying concludes. Notably, both non-motile and motile chlamys droplets exhibit a ring width of $0.17-22$ mm [indicated by the yellow arrow in Figure~\ref{fig6}(IV-V)]. This stage provides a distinctive fingerprint-like pattern [Figure~\ref{fig6}(IV-V)], highlighting the differences between patterns arising from motile and non-motile chlamys droplets in the presence of localized nutrients.

\subsubsection{Chlamys droplets in the absence of local nutrients}
\label{subsec:phyDI}

The drying progression and pattern formation of motile and non-motile chlamys in the absence of local nutrients are depicted in Figure~\ref{fig7}(I-VII). In Figure~\ref{fig7}(I)a-c, most chlamys are transported towards the droplet's periphery via capillary flow. The inward motion of these transported chlamys is similarly observed (Figure~\ref{fig7}(I)d-f). A new phenomenon emerges for motile chlamys without nutrients in the DDP stage, with the formation of diverse clusters while ample water remains in the droplet. Neighboring chlamys converge, creating small clusters that subsequently merge into larger ones. As time elapses (Figure~\ref{fig7}(I)g-l), these clusters also exhibit rotational motion. Cluster counts are performed, as highlighted in the zoomed image (Figure~\ref{fig7}(I)j) with boxed red labels. Smaller clusters are marked in blue (Figure~\ref{fig7}(II)), housing 5-7 chlamys each. A histogram in Figure~\ref{fig7}(III) illustrates the distribution of chlamys numbers per cluster, varying from ${\sim}13$ to ${\sim}59$, averaging around $34$. Individual cluster instances are shown in Figure S3 of the supplementary section.

\begin{table}[h]
\centering
\caption{Summary of different events occurs in the drying droplets of motile and non-motile chlamys with and without local nutrients; where different stages include Droplet Deposition (DD), Capillary Flow (CF), Dynamic Droplet Phase (DDP), Aggregation Phase (AP), and Dried Morphology (DM).}
\label{tab:my-table}
\resizebox{\columnwidth}{!}{%
\begin{tabular}{llllll}
\hline
Stages & Characteristics & Motile  & Non-motile  & Motile & Non-motile  \\
       &                 & + without nutrients  & + without nutrients & + with nutrients  & + with nutrients               \\ \hline \\
DD  & Spherical-cap shape & YES                            & YES                & YES  & YES               \\ 
    & (dark \& gray texture) &                    & & & \\ \hline \\
CF  & Different flows     & YES  (faster move)                              & YES   (slower move)         & YES  (faster move)   & YES   (slower move)  \\ \hline \\
DDP & Coffee-ring         & YES                                 & NO (uniform deposition) & YES & YES                     \\ 
    & Cluster formation   & YES                                 & NO                       & NO  & NO                      \\ \hline \\
AP  & Aggregation         & YES  & YES                      & YES & YES                     \\ 
    & Crack formation     & NO                                  & NO                       & NO  & NO                      \\ 
    & (due to mechanical stress) & & &  &  \\ \hline \\
DM  & Droplet pinning     & YES                                 & YES                      & YES & YES                     \\ 
    & Unique patterns     & YES                                 & YES                      & YES & YES                     \\ 
    & Reverse coffee-ring   & NO                                 & NO                      & NO & YES                     \\ 
    & (droplet edge with homogeneous texture) & & &  &  \\     \hline \\
\end{tabular}%
}
\end{table}

The mechanism behind cluster formation is exhibited in Figure~\ref{fig7}(IV). As drying progresses, chlamys begin interacting and assembling into clusters. Attractive hydrodynamic forces primarily drive this aggregation \cite{hokmabad2022spontaneously}. The cluster formation initially resembles chains of 3-4 chlamys, followed by ordered arrangements of 6-10 chlamys. These clusters merge to form larger entities, exhibiting cooperative rotation. While this ordered configuration isn't sustained for a longer time, cluster formation remains consistent till the significant water is evaporated in the AP stage. To ascertain the role of drying, sealed chamber experiments confirm this cluster formation (Figure S4 in supplementary section), emphasizing the significance of nutrient absence. The lack of nutrients seemingly induces local stress on motile chlamys, prompting cooperative aggregation into larger colonies, possibly as a response to inadequate nutrient supply. The unity among chlamys could be attributed to the competition against stress in nutrient-deficient conditions. A similar coffee-ring width of ${\sim}0.2$ mm is also found, irrespective of the availability of the nutrients. However, we do not find any trend when the cluster diameter and the number of chlamys in each cluster are plotted as a function of the radial distance (distance measured from the center). Even the cluster diameter normalized by the corresponding ``coffee ring width" as a function of the radial distance does not show any trend (see Figures S5-S6 in the supplementary section).

In contrast, non-motile chlamys without nutrients do not exhibit significant coffee-ring patterns (Figure~\ref{fig7}(VI-VII)). Figure~\ref{fig7}(VI)a-l shows the time evolution of non-motile chlamys, uniformly depositing compared to other conditions. The morphological outcome post-drying is displayed in Figure~\ref{fig7}(VII). Specifically, in the non-motile case without nutrients, the outermost region demonstrates a smooth, uniform texture, followed by the chlamys.

Addressing the persistent pinning of droplets across classes (motile vs. non-motile with and without nutrients), we conducted experiments using (i) centrifuged deionized water carrying the non-motile chlamys and (ii) pure deionized water. The resulting drying evolutions are shown in Figure S7(I-II) of the supplementary section. Observations indicate fine deposits in the former case, implying that residual TAP medium particles contribute to droplet pinning rather than chlamys themselves. Table~\ref{tab:my-table} tabulates different events occurring during the drying process of chlamys of different classes, i.e., non-motile + with nutrients,  motile + with nutrients, non-motile + without nutrients, and motile + without nutrients.

\subsection{Quantitative textural image statistics of chlamys droplets}
\label{subsec:tex}

\begin{figure*}[h]
\centering
  \includegraphics[height=8cm]{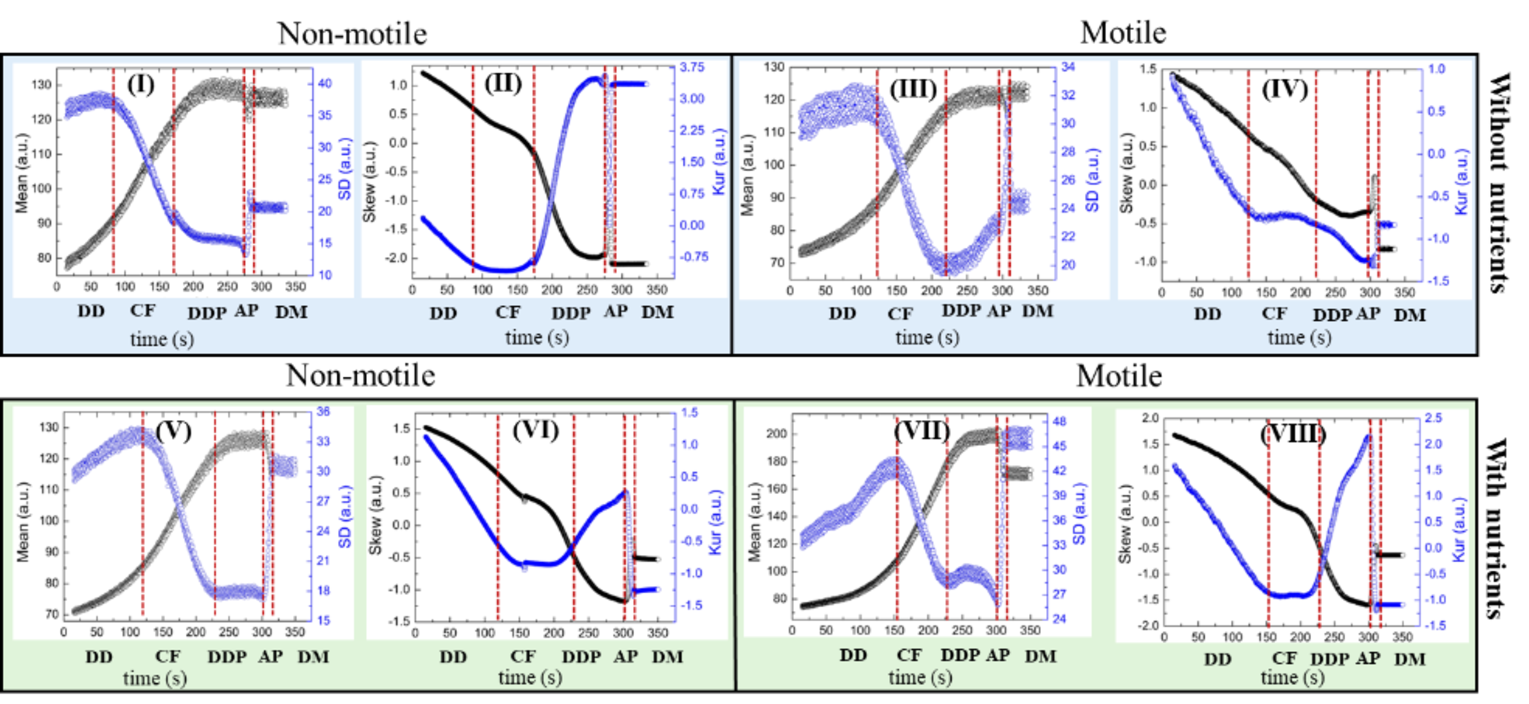}
  \caption{(I-VIII) Drying evolution of the statistical textural image parameters, the Mean, standard deviation (SD), skewness (Skew) and kurtosis (Kur) [in arbitrary units, (a.u.)] of motile and non-motile chlamys droplets with and without local nutrients. Mean and Skew are depicted by black lines, whereas SD and Kur are by blue lines. Different stages are highlighted by red lines: Droplet Deposition (DD), Capillary Flow (CF), Dynamic Droplet Phase (DDP), Aggregation Phase (AP), and Dried Morphology (DM). The x-axis represents time (in seconds) during the drying process.}
  \label{fig1}
\end{figure*}

Figure~\ref{fig1}(I-VIII) illustrates the drying evolution of the textural parameters, including mean, standard deviation (SD), skewness (Skew), and kurtosis (Kur) for motile and non-motile chlamys droplets, with and without local nutrients. These parameters capture various statistical features within the droplet area. 

The mean value initially experiences a slow increase during the droplet deposition (DD) stage, followed by a rapid rise in the capillary flow (CF) stage, with a variation of ${\sim}15$ a.u. and ${\sim}50$ a.u., respectively. Subsequently, during the dynamic droplet phase (DDP), there is minimal change or fluctuation in the mean value. A dip occurs in the aggregation (AP) stage, leading to a decrease of ${\sim}15$ a.u. The mean value increases again to reach saturated values in the final dried morphology (DM) stage. The standard deviation (SD) ranges from ${\sim}15$ a.u. to ${\sim}45$ a.u. The general trend is a slow increase in the DD stage, followed by a rapid decrease in the CF stage. However, the DDP stage exhibits distinct characteristics for each type of chlamys droplet. For instance, in the presence of nutrients, motile chlamys droplets display a hump-like feature during the DDP stage, while a slow increase is observed in their absence. Non-motile chlamys droplets show minimal fluctuations in the presence of nutrients, whereas a slight decrease is observed without nutrients. The AP stage is the shortest phase, characterized by a rapid rise followed by a decrease until saturation is reached in the DM stage. Notably, during the DDP stage, the SD captures different details while the mean remains relatively constant. During the AP, the complexity intensifies, resulting in a sharp increase in the SD values [see Figure~\ref{fig1}(I, III, V, and VIII)].

The skewness (Skew) exhibits a decreasing trend until the DDP stage, after which it increases during the AP stage. Following this, the values stabilize, similar to the mean and SD values. While the Skew does not display a distinct pattern with respect to motility and nutrient availability, the kurtosis (Kur) exhibits notable trends during the DDP and AP stages. In the DDP stage, the Kur exhibits a linear trend, with adjusted R$^2$ values of 0.89-0.96, although the slope values differ. The slope for motile chlamys droplets with and without nutrients is ${\sim}0.03s^{-1}$ and $-0.006s^{-1}$, respectively. In contrast, the slope values for non-motile chlamys droplets with and without nutrients are ${\sim}0.01s^{-1}$ and ${\sim}0.04s^{-1}$, respectively. In the AP stage, the Kur decreases in the presence of nutrients, regardless of motility. However, in the absence of nutrients, it increases for motile chlamys, while it alternately increases and decreases for non-motile chlamys droplets [see Figure~\ref{fig1}(II, IV, VI, and VIII)].

\begin{figure*}[h]
\centering
  \includegraphics[height=13cm]{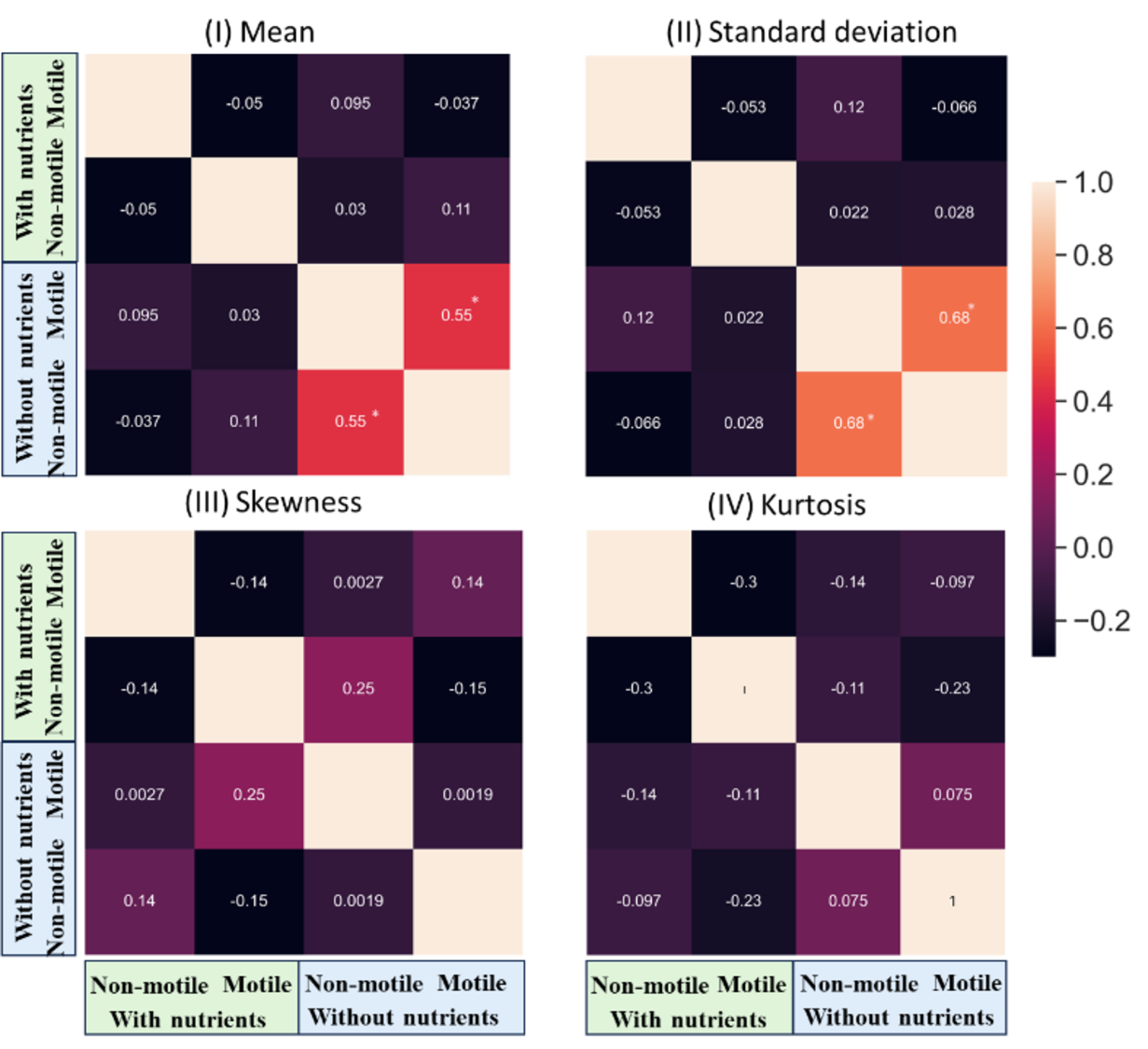}
  \caption{Correlation matrix of non-motile and motile chlamys with and without local nutrients for different textural image statistical parameters: (I) Mean, (II) Standard deviation, (III) Skewness, and (IV) Kurtosis. These show the correlation coefficients (between -1 and 1), measuring the strength of the relationship between the different combinations. This correlation is performed on the data acquired in the final stage (Droplet Morphology, DM) of the drying process. The asterisk (*) mark shows the highly correlated and significant combinations.}
  \label{fig8}
\end{figure*}

Visual examination of the DM stage, particularly in Figure~\ref{fig1a}, reveals that the patterns are very distinct, suggesting a minimal correlation between patterns of non-motile and motile chlamys with and without nutrients. To assess whether all the textural statistical parameters (mean, standard deviation (SD), kurtosis, and skewness) support this visual observation, we computed the correlation matrix at the DM stage [see Figure~\ref{fig8}(I-IV)]. Remarkably, skewness and kurtosis exhibit correlation values lower than ${\sim}0.25$ across all classes (motile+with nutrients, motile+without nutrients, non-motile+with nutrients, and non-motile+without nutrients). In contrast, mean and SD parameters for non-motile and motile chlamys in the absence of local nutrients display higher correlations of 0.5-0.7, implying that their morphological patterns are not significantly distinct. This is possible due to the fact the ranges of Mean and SD are quite similar (with mean $\sim 115-130$~a.u., and SD $\sim 20-25$~a.u.) for motile and non-motile chlamys without local nutrients [see Figure~\ref{fig1}(I, and III)]. 

An intriguing observation arises when examining the skewness vs. kurtosis plot (see Figure S8(I-II) in the supplementary section). Skew and Kur offer valuable insights into data distribution and shape. In the presence of nutrients, there is symmetry and balance in the data. Conversely, the absence of such a clear pattern in skewness-kurtosis relationships when nutrients are absent suggests a more complex and diverse distribution (see Figure S8(I) in the supplementary section). Notably, from Figure~\ref{fig1}, the AP and DM stages exhibit a chaotic appearance, indicating higher variability and unpredictability in chlamys droplet behaviors during these stages. Indeed, the skewness-kurtosis plot can unveil interesting insights into system dynamics. For instance, considering all stages, scattered data points along negatively skewed values suggest greater diversity or complexity in behavior. However, focusing on the first three stages (DD, CF, and DDP) of chlamys drying reveals a specific trend (symmetry) in the plot (Figure S8(II) in the supplementary section). The alignment of data points within a specific region indicates consistent behavior or arrangement of chlamys particles during those stages. This symmetry persists until the DDP stage, after which it breaks as the drying process advances. Deciphering subtle changes in statistical parameters like skewness and kurtosis across different drying stages can offer crucial insights into underlying mechanisms and dynamics. Such information could potentially guide the development of refined models or theories, aiding the explanation of observed behaviors and guiding further experimental exploration.

Therefore, the analysis of these textural parameters not only helps in quantifying the evolving stages of the drying process of chlamys droplets but also reveals the distinctive behaviors between motile and non-motile chlamys in the presence or absence of local nutrients. The variations observed in mean, standard deviation, skewness, and kurtosis provide valuable insights into the intricate dynamics occurring during the drying process, including aggregation, flow patterns, and particle distribution. The distinct trends and patterns observed in these parameters during different stages of drying provide a robust basis for training models that can potentially automate the classification process and provide deeper insights into the complex behavior of chlamys droplets under varying conditions. However, it is crucial to note that not all textural parameters are equally effective in implementing supervised machine learning algorithms. This underscores the importance of careful consideration and selection in model development, with skewness and kurtosis being particularly useful in this study among the four parameters analyzed in the dried morphology stage.

In conclusion, this study illuminates the multifaceted potential of \textit{Chlamydomonas reinhardtii} (chlamys) as an exceptional algal model within the context of sessile drying droplets cross-linking the micro and macro length scales. The interplay between motility-nutrient interactions, localized nutrient scarcity, and mechanical stress during drying has been unraveled through different image analyses on chlamys. Identifying five distinct drying stages-- Droplet Deposition, Capillary Flow, Dynamic Droplet Phase, Aggregation Phase, and Dried Morphology-- provides a comprehensive understanding of the intricate processes underlying drying dynamics. Notably, mechanical stress is found to be ineffective in inducing cracks. On the other hand, local nutrient deficiency induces stress within chlamys, fostering cooperative aggregation, unity, and cluster formation. The intriguing observation of a reverse coffee-ring effect in the presence of local nutrients introduces complexity to these findings. This droplet drying-based tool opens promising avenues for the bio-behavioral responses to external stimuli, and the protocol can be used for different classification purposes. Notably, the insights gained from this study hold applicability beyond chlamys, extending to various microorganisms, not only contributing to the understanding of the drying process but also catalyzing advancements in diverse fields spanning biology, biomaterials, and health sciences.

\section{Methods}
\label{sec:exp}
\subsection{Cell culture, samples, and its preparations}
Chlamys, a type of eukaryotic microorganism, boasts dimensions of ${\sim}10~\mu$m in length and possesses flagella measuring around $10~\mu$m \cite{jeanneret2016brief}. Notably, the algal strain CC-124 features an eyespot—an organelle sensitive to light—enabling phototaxis. Although a flagellum's primary role involves propelling cellular motion, it also serves as a sensory apparatus, responding to chemical, mechanical, and environmental cues such as temperature and humidity. Intriguingly, CC-124 demonstrates negative phototaxis, signifying a tendency to move away from light sources.

The chlamys strain CC124 cultures were cultivated axenically using a Tris-Acetate-Phosphate (TAP) medium \cite{goldschmidt1989preparation} at $21^{\circ}$C under continuous fluorescent illumination (100 $\mu$ E m$^{-2}$ s$^{-1}$, OSRAM Fluora). The light-dark cycle followed a 12-hour cycle each. The TAP medium composition comprises Tris base, TAP Salts Stock Solution, Hutner's Trace Metal Solution, Glacial or anhydrous acetic acid, and phosphates, which facilitate cell synchronization. Harvesting was done at $7-10 \times 10^6$ cells mL$^{-1}$ during exponential growth, and daily transfers were made to fresh phosphate-buffered medium. The growth rate was monitored daily, and only healthy cells were used in this experiment. The growth curve is illustrated in Figure S9(I), and cell motility is depicted in S9(II) of the supplementary section. Two conditions were examined: one with optimal nutrient availability and the other without nutrients, i.e., de-ionized water (DI). For the experiment with DI, sub-sampled cells were subjected to centrifugation at ${\sim}800$~r.p.m. for ${\sim}10$~minutes \cite{jeanneret2016entrainment}. The supernatant was then exchanged with deionized water (DI) from Millipore, with a resistance of $18.2$M$\Omega$.cm. To induce cell immobility, ${\sim}1$~$\mu$L of Iodine solution, acting as a red dye, was introduced into the eppendorf tubes containing 1.5~mL of the algal solution. This mixture was left undisturbed for roughly ${\sim}10$~minutes. Experiments involving motile cells were executed within seconds after sample preparation, mitigating the possibility of any cell lysis (if present) before the drying phase.

${\sim}1$~$\mu$L of each prepared sample was pipetted onto a microscopic coverslip (Catalog number 48366-045, VWR, USA), forming a circular droplet with a radius of ${\sim}1$~$mm$. All experiments were conducted under ambient conditions, encompassing a room temperature of ${\sim}25^{\circ}$C and a relative humidity of ${\sim}50$~\%. To ensure reliability, each experiment was replicated twice. The morphological dried patterns of each chlamys are illustrated in Figure S10 of the supplementary section.

\subsection{Image acquisition}
The progression of droplet drying was captured at a rate of 4 frames per second (fps) utilizing a FLIR camera (Catalog no KCC-REM-PGR-GS41) mounted on a Nikon Eclipse E200 microscope, employing bright-field illumination. To prevent any phototactic response from the cells, a long-pass filter with a cut-off wavelength of 765 nm was integrated into the optical pathway. Time recording was initiated upon droplet deposition onto the coverslip. Images were captured at a resolution of $1600\times1600$ pixels. To maintain minimal fluctuations in background (coverslip) intensity, the lamp intensity remained constant throughout the drying process. Prior to image processing, pixel values were converted to real-space lengths using a calibration slide. All images were transformed into 8-bit grayscale for enhanced clarity.

The temporal evolution of the contact angle of the chlamys solution during drying was assessed using the drop shape analyzer (DSA 100, KR$\ddot{U}$SS, GmbH). Normalized with respect to the initial contact angle, the normalized contact angle, and the height of the droplet apex in relation to drying time is depicted in Figure S1 of the supplementary section.

\subsection{Image processing}
We employed the \textit{oval tool} within ImageJ \cite{abramoff2004image} to define a circular region of interest (ROI) on the image. Notably, the 8-bit image encompasses gray values spanning from 0 to 255. Our assessment revolved around four essential textural first-order statistics (FOS), namely, the Mean, standard deviation (SD), kurtosis (Kur), and skewness (Skew). These FOS parameters were then extracted from the 8-bit gray images. Subsequently, these textural parameters were harnessed as a feature vector. This vector was employed within a machine learning framework to predict various combinations, thus enabling insightful outcomes.

\section*{Author contributions statement}
Conceived and Designed: A.P. and M.Y., Data collection and Compilation: A.P., Data Analysis, and Interpretation: A.P., A.S., and M.Y., with A. S., and M.Y. supporting in specific analyses, Conclusions: A.P, A.S., and M.Y., Initial Draft: A.P., Revision, and Final editing: A.P., A.S., and M.Y., and Overall Supervision: A.P.

\section*{Acknowledgments}

This research is supported by the Department of Physics at the University of Warwick, UK, and the Graduate School of Arts and Sciences at The University of Tokyo, Japan. The authors would like to express their gratitude to M. Polin, Associate Professor at the University of Warwick, for his valuable support. Financial backing for this study is provided by the Leverhulme Trust (Grant No. RPG-2018-345). Thanks to the Japan Society for Promotion of Science (JSPS), KAKENHI Grant No. 23KF0104, for supporting this work. A. Pal expresses appreciation for the JSPS International Postdoctoral Fellowship for Research in Japan (Standard) for the period 2023-25. A. Sengupta thanks the Luxembourg National Research Fund's ATTRACT Investigator Grant (Grant no.~A17/MS/11572821/MBRACE) and CORE Grant (C19/MS/13719464/TOPOFLUME/Sengupta) for supporting this work.

\section*{Conflict of interest}
The authors declare no conflict of interest.

\section*{Data Availability Statement}
The data that support the findings of this study are available from the corresponding author upon reasonable request.

\bibliography{main_screports}

\end{document}